\documentstyle[11pt,epsfig,subfigure]{article}

\title {\Huge\bf On the observability of free quarks 
       near their production thresholds \vspace {20 mm}}
\author{
 A. Markou\\
\small\rm INSTITUTE OF NUCLEAR PHYSICS\\
N.C.S.R Demokritos,\\
 15310 Aghia Paraskevi, Greece \\
\small\rm a.markou@cern.ch}
\vspace{1cm}
\date{October  5,1994}
\vspace{6cm}
\begin{document}
\pagestyle{empty}
\maketitle
\pagestyle{empty}
\vspace{4cm}
\begin{abstract}
\thispagestyle{empty}
\normalsize\rm
It is proposed to search for free quarks near their production thresholds.
If the current-quark masses for u, d quarks are smaller than the pion mass,
$u\bar{u}$ and $d\bar{d}$ pairs produced below the pion pair threshold, would not have
sufficient energy to hadronize and could be observable. In the case of large
u,d quark masses and also for the heavy quarks, production near threshold
 may show hadronization probability $<1$, and lead to
free quarks.  Searches for fractional charges in stable matter could be more
promising, through irradiation of samples at energies near the quark pair production
thresholds.\\

\vspace{1cm}

\end{abstract}
\large\rm
\vspace{-22cm}\hspace{12.5cm}DEMO 94/27
\newpage
\section { Introduction }
\pagestyle{plain}
\pagenumbering{arabic}

Numerous free quark searches have been conducted up to now [1],[2],[3],[4]. 
The results have been negative, with the exception of one experiment [3]
 which claimed to have observed fractional charges in stable matter, 
 but which was contradicted by other experiments. A commonly accepted 
 explanation for the non-observation of free quarks, is that the quarks 
 are confined within hadrons. Evidence that this confinement conjecture 
 is valid, was given from numerical calculations in the framework of 
 lattice QCD [5], but these calculations allow deconfinement under 
 certain conditions, and since a rigorous analytic proof does not exist,
 the question of the existence of free quarks is still open and
the search for free quarks is going on [4].

The non observation of free quarks could also be explained, 
if the cross-section of the quark interaction with matter is very
 high as has been suggested in [6]. In fact only few of the quark 
 search experiments have used little material ($< 1$\% inter length) between 
 the interaction region and the detection apparatus [2]. These 
 experiments further improved the experimental upper 
 limits for the production of free quarks to
 $R=\sigma (qq)/ \sigma (\mu^+ \mu^-)=10^{-4}$, 
 and it is now believed that any produced quarks hadronize 
with probability  1.

It was expected that with increasing accerelator energies, 
an energy could be reached where free quarks may exist.
 Increasing of the energy though, may simply 
result in further increasing the hadron multiplicity according 
to the relation
$<n>=A+B\times lns+C\times(lns)^2$
 measured up to 
existing energies.

\section{ Implications of small quark-masses}

A possibility for free quark observation may exist, especially 
for the light quarks, provided that their masses are small. 
Mass calculations from different sources, as the QCD sum rules,
 SU(4) symmetry and grand unified theories, give few 
$ MeV/c^2$ masses for the u and d "current quarks" 
as is discussed in the review paper [7] and also later [8]. 
Calculations in the framework of lattice QCD [9], 
are consistent with the above mentioned values.
Models based on extended Technicolor produce u and d quark 
masses in the region 35 to 70 $MeV/c^2$ [10], whereas calculations 
in a six-dirnentional SO(12) theory [11] argues that 
$m_{/mu}\gg m_u,m_d,m_e$.
In QCD, the used masses are the running masses, 
defined at energy scale $\mu = 1 GeV$ and they are expected to 
grow with smaller scales in a way similar to the running of 
the QCD coupling constant. However, in some cases (see for 
example [12]), the masses are considered as electromagnetic 
self energies and the color force is not expected to contribute 
substantially to them. In such a case the growing of the running 
masses with smaller scales would be limited.

Therefore, and despite the fact that according to nonrelativistic 
potential models the masses of the u and d quarks bound in hadrons 
(constituent quarks, considered to contain a valence quark and many
 qq pairs and gluons), are about 300 $MeV/c^2$, we accept the existence 
 of evidence from a variety of theoretical calculations, that the 
 masses of the u and d quarks -if they would exist as free particles- 
 should be considerably smaller than the pion mass. A mass of the 
 u or d quarks smaller than the pion mass, could allow deconfinement 
 of these quarks at low energies:\\
 
 \pagebreak
In $e^+e^-$ collisions at C.M. energies below the pion pair
 production threshold, $u\bar{u}$ or $d\bar{d}$
  pairs produced electromagnetically 
 according to the diagramm shown on Figure 1, would lack energy to
  produce the lowest hadrons i.e they could not hadronize (unless
   hadrons less heavy than pions would exist), and could be observable.

\begin{figure}
\begin{center}
\mbox{\epsfig{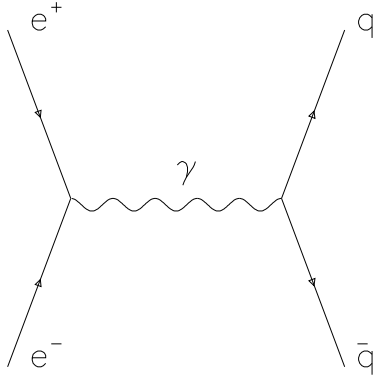}}
\caption{$q\bar{q}$ production in low energy $e^+e^-$ one photon annihilation}
\end{center}
\end{figure}

Along the same line of thinking, if the masses of the u and d 
quarks are larger than the pion mass, the small available energy 
for hadronization (in the case of production near threshold) may 
lead to a hadronization probability smaller than 1, which has 
been observed up to now at high energies. Therefore free quarks may 
be observable near their production thresholds.

 \section{Free quark observability}

To estimate the expected $q\bar{q}$ rates and compare with possible
 backgrounds, we use the following expression for the 
 cross-section [13] calculated exactly in the one photon 
 annihilation of QED. which describes the production of 
 fermion-antifermion pairs in $e^+e^-$ interactions 
            
               \begin{equation}
              \sigma=e_f^2N_c \frac{\pi \alpha^2(\hbar c)^2}{3E_b^2}
              \sqrt{1-m^2/E_b^2}(1+m^2/2E_b^2)
               \end{equation}

where $e_f$ is the quark charge, $N_c$ is a factor for the 
quark color and $E_b$, is the beam energy. This expression 
is obtained by integrating the corresponding differential cross section:

\begin{equation}
         \frac{d\sigma}{d\Omega}=e_f^2\frac{\alpha^2N_c(\hbar c)^2}{4s}
       \beta (1+cos^2\theta +(1-\beta^2)sin^2\theta)  
               \end{equation}

\pagebreak
Fig. 2 shows the cross section for the production of 
$u\bar{u}$ and $d\bar{d}$ pairs in $e^+e^-$ collisions, calculated using 
(1), with masses $m_u=50$  and 100 $MeV/c^2$ and $m_d$=l00 $MeV/c^2$.

\begin{figure}[h]
\begin{center}
\mbox{\epsfig{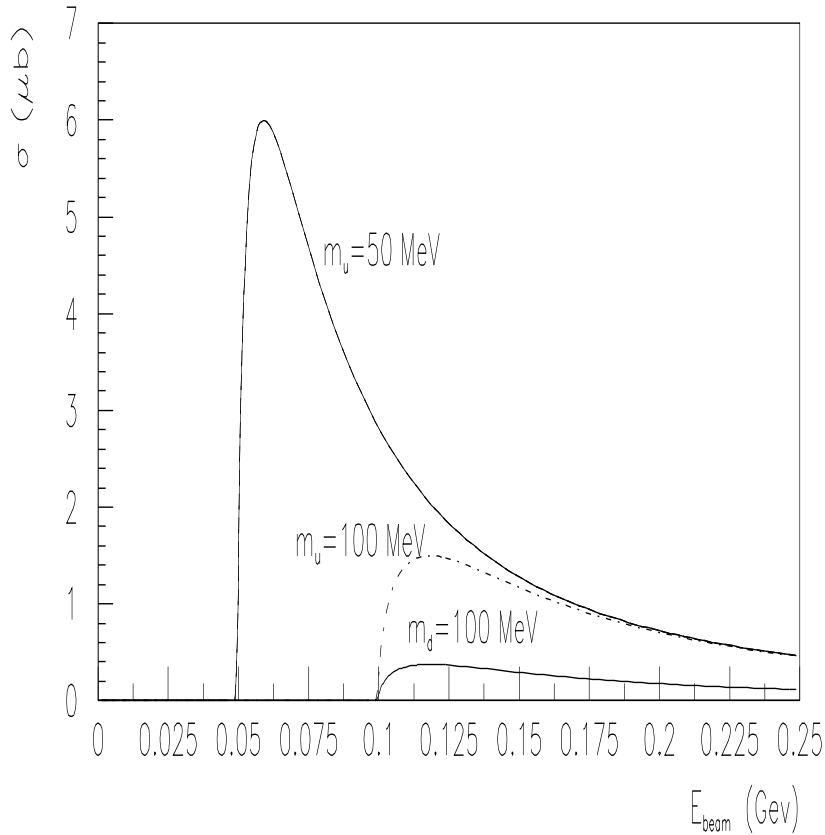}}
\caption{Calculated total $e^+e^- \rightarrow q\bar{q}$ cross section in 
the low energy region.}
\end{center}
\end{figure}

The u and d quarks produced in $e^+e^-$ collisions as 
described above, could be detected and their charge and 
mass could he calculated from the measured momenta, using 
the energy conservation criterion, provided that the Bhabha 
scattering background would not he too large. Additional 
measurements (time of flight, dE/dx) will help quark 
identification. To check if the background is not too large, 
we have calculated $sd\sigma /d\Omega$ for $e^+e^- \rightarrow u\bar{u}$
and $d\bar{d}$ using relation (2), 
for $\beta=l$. The results are ploted on Fig. 3. Together with a 
$e^+e^- \rightarrow e^+e^-$
calculation using the expression:\\

\pagebreak

\begin{equation}
          s \frac{d\sigma}{d\Omega}= \frac{\alpha^2(\hbar c)^2}{4}
          \frac{(3+cos^2\theta)^2 }{(1-cos\theta)^2}
               \end{equation}

It is clear that for $cos(\theta) < 0$, $u\bar{u}$ and $d\bar{d}$ 
production rates 
are not too far below Bhabha scattering rates. If no free u,d 
quarks would he observed under these conditions, lower limits 
can be set on their masses. It is interesting to note that no 
free quark search has been conducted up to now in this energy 
region.

\begin{figure}[h]
\begin{center}
\mbox{\epsfig{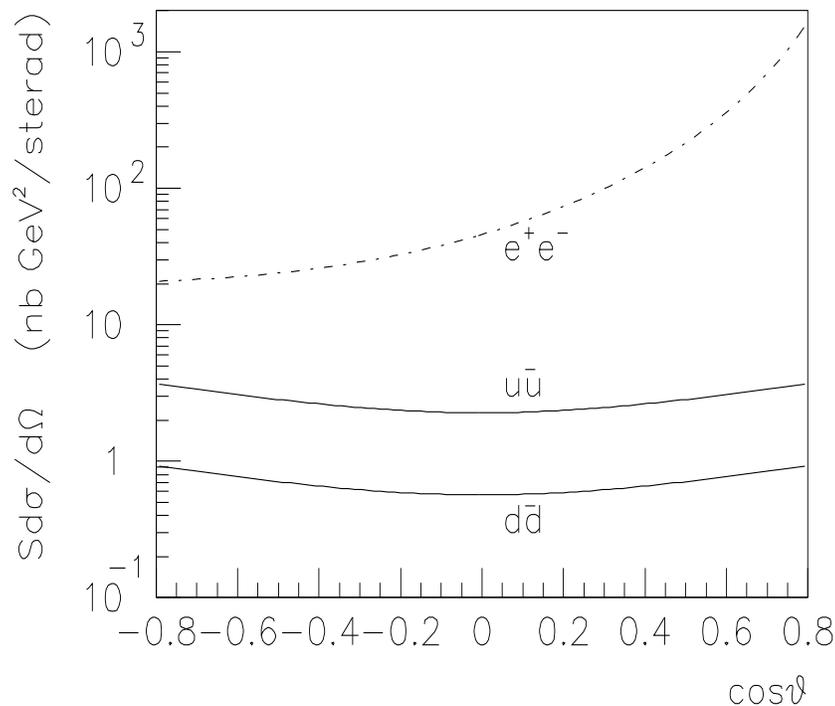}}
\caption{Calculated differentiaJ cross section for 
the production of quark pairs in low energy $e^+e^-$ 
collisions compared to the Bhabha cross section.}
\end{center}
\end{figure}

In an analogous way,
 free s quarks may be observed at C.M. energies 
 below the kaon threshold, because $130< m_s<230 MeV/c^2$ [7].
  Below this threshold the produced strange quarks would 
  lack energy to hadronize to Kaons (unless hadrons with 
  strange quark content and less heavy than Kaons existed). 
  Similar argumentss are valid for the heavy quarks.

The detection of free quarks may be also possible in other 
reactions -as for example in photoproduction- below the 
appropriate thresholds. 
In photoproduction, the ratio 
$\sigma(q\bar{q}/\sigma(e^+e^-)\sim10^{-4}  $
for $m_u"=10MeV$ and $\sim 10^{-6} $
for $m_u=100MeV$
 because of the dependence of the cross section on $(e_f)^4/m^2$.
Therefore the electron-positron background would be large. It is possible 
to discriminate against $e^+e^-$ at large production angles, 
because $e^+e^-$
 are
 produced mainly in the beam direction. In a magnetic spectrometer, quarks 
 produced at large angles to the beam can be separated from 
 $e^+e^-$ pairs through 
 the energy conservation requirement plus dE/dx and time of flight cuts.
  In this case too.
the alleged large quark interaction cross section with matter may 
dictate the use of thin targets and evacuated spectrometers. Free 
quark searches carried out in photoproduction so far, gave negative 
results but have been performed at energies $>6$ GeV [14] and used 
large amounts of material in front of the detectors.

For similar reasons, free quark search in stable matter, may be
 more promising than up to now, if the samples under investigation 
 for fractional charges would be previously irradiated at energies 
 near the quark pair production thresholds.


\begin{thebibliography}{99}

\bibitem{el:chain1}
 L. Jones, Rev. mod. Phys. 49 (1977) 717, and references therein.\\
L. Lyons, Prog. Part. Nucl. Phys. 7 (1981) L57, and references therein.\\
JADE Collab., P. Shacht, Leipzig Conf. 1984.\\
J. E. Hulth, Ph.D Thesis LBL-18341/1984, unpublished.\\
L. Lyons, Phys. Rep. 129 (1985) 225, and references therein.\\

\bibitem{el:chain2} JADE Collab., W. Cartel et aL, Z. Phys. C 6 (1980) 295.\\
W. Guryn et al., Phys. Lett. 139 B (1984) 313.\\
ARGUS Collab., H Albrecht et aL, Phys. Lett. 156 B (1985) 134.\\

\bibitem{el:chain3} G. S. LaRue et aL, Phys. Rev. Lett. 38 (1977) 1001.\\
G. S. LaRue et aL. Phys. Rev. Lett. 42 (1979) 142 and 2166.\\
G. S. LaRue et aL, Phys. Rev. Lett. 46 (1981) 967.\\

\bibitem{el:chain4} W. Innes, M. Pei-l, J. Price, SLAC-PUU 3867/1986.\\
BEBC results, D. AHasiaet aL, Phys. Rev. D 37 (1988) 219.\\
H.S Matis et aL, Phys. Rev. D 39 (1989) 1851.\\
CLEO Collab., T. Bowcock et aL, Phys. Rev. D 40 (1989) 263.\\
P.F. Smith, Ann. Rev. NucL Part. Sci. 39 (1989) 73, and references therein.\\
Kamiokande II Collab., M. Mori et aL, Phys. Rev. D 43 (1991) 2843.\\
H. S. Matis et aL, NucL Phys. A 525 (1991) 513.\\
M. Basile et al, Nuov. Cim. A 104 (1991) 405.\\
CDF Collab., F. Abe et aL, Phys. Rev. D 46 (1992) R1889.\\
G. L. Homer et al, Zeit. Phys. C 55 (1992) 549.\\
ALEPH Collab., D. Buskulic et aL, Phys. Lett. B 303 (1993) 198.\\
A. G. U. Perera et al., Phys. Rev. Lett. 70 (1993) 1053.\\
C. Hendricks, K. Lackner, M. Peri, G. Shaw, SLAC-PUB-6288/1993.\\

\bibitem{el:chain5} M Creutz, Phys. Rev. D 21 (1980) 2308.\\
M. Creutz, K.J.M. Moriarty, Phys. rev. D -26 (1982) 2166.\\
D. Barkai, et al, Phys. Rev. D 29 (1984) 1207.\\
E. Shuryak, Phys. Rep. 89 (1984) 163.\\

\bibitem{el:chain6}  A. De Rujula et al.,Phys. Rev. D 17 (1978) 285.\\
J. Orear Phys. Rev. D 18 (1978) 3504.\\
D. Garelick. Phys. Rev. D 19 (1979) 1026.\\

\bibitem{el:chain7}  J. Gasser, H. Leutwyler, Phys. Rep. 87 (1982) 77 and references therein.\\

\bibitem{el:chain8}  A.L. Kataev et aL Phys. Lett. 123 B (1983) 93.\\
V.P. Efrosinin, D.A. Zaikin, Yad. Phys. 37 (1983) 1532.\\
M.K. Volkov et al. Ya.d. Phys. 39 (1984) 924.\\
V. P. Efrosinin et al., Z. Phys. C 28 (1985) 211.\\
D. Ebert, Z. Phys. C28(1985)433.\\

\bibitem{el:chain9}  A. Hasenfrantz et al.. CERN TH. 3220.\\
F. Fucito et al., Nnc!. Phys. B 210 (1982) 407.\\

\bibitem{el:chain10}  G. Aubrecht. Phys. Lett. 117 B (1982) 342.\\
H. Sugavara, Phys. Rev. D 30 (1985) 2396.\\

\bibitem{el:chain11}  C. Wetterich, Nucl. Phys. B 261 (1985) 461.\\
\bibitem{el:chain12}  U. Baur, H. Fritzsch. Phys. Lett. 134 B (1984) 105.\\
\bibitem{el:chain13}  F. Close, An introduction to quarks and partons, Academic Press 1979.\\
\bibitem{el:chain14}  G. Bathow et al., Phys. Lett. 25 B (1967) 163.\\
J. Foss et al., Phys. Lett. 25 B (1967) 166.\\
F. Bellamy et al., Phys. Rev. 166 (1968) 1391.\\

\end{thebibliography}
\end{document}